\documentclass[prb,twocolumn,preprintnumbers,amsmath,amssymb]{revtex4}
\usepackage{graphicx}

\begin{document}

\title{ \bf  Elementary excitation in a supersolid   }
\author{ \bf  Jinwu Ye  }
\affiliation{ Department of Physics, The Pennsylvania State
University, University Park, PA, 16802 }
\date{\today}

\begin{abstract}
  We study  elementary low energy excitations inside a
  supersolid. We find that the coupling between the longitudinal lattice
  vibration mode and the superfluid mode leads to two longitudinal
  modes ( one upper  branch and one lower branch ) inside the supersolid,
  while the transverse modes in the supersolid stay the same as those inside a normal solid. We also work out
  various experimental signatures of these novel elementary excitations
  by evaluating the Debye-Waller factor, density-density correlation, vortex
  loop-vertex loop interactions, specific heat and  excess entropy from the vacancies per mole.
\end{abstract}


\maketitle

{\bf 1. Introduction.}
     A supersolid is a state which has both a solid order and a
     superfluid order. The possibility of a supersolid phase in $^{4}He$ was
     theoretically speculated in 1970 \cite{and}.  Over the last 35
     years, a number of experiments have been designed to search for
     the supersolid state without success. However, recently, by
     using torsional oscillator measurement, a PSU group lead by Chan observed
     a marked $ 1 \sim 2 \% $  NCRI of solid $^{4}He$
     at $ \sim 0.2 K $ in bulk $^{4}He$ \cite{science}.
     The authors suggested that the supersolid state of $^{4}He$ maybe responsible
     for the NCRI. The PSU experiments rekindled extensive
     both theoretical \cite{micro,macro,dor,ander,qglprl,long} and
     experimental \cite{annealing,massflow,melt} interests.
     The torsional oscillator measurement
     is essentially a dynamic measurement, as suggested in \cite{micro},
     other possibilities can also lead to the NCRI  observed in the Kim-Chan
     experiments. Obviously, many other thermodynamic
     and equilibrium measurements are needed to make a definite conclusion.
     Some interesting physics near the {\em finite } temperature
     normal solid to supersolid transition was explored in
     \cite{dor}. The fundamental questions to be addressed in this paper
     is what are the elementary excitations in a supersolid at zero and very low temperature and what
     are the physical consequences of these elementary excitations
     which can be probed by various experimental techniques such as
     X-ray scattering, neutron scattering, acoustic wave attenuation and heat capacity in solid Helium 4.
     In principle, if these elementary low energy excitations can be detected by these experiments  can
     prove or disprove the existence of the supersolid in Helium 4.


{\bf 2. Elementary excitations in a SS.}
   Classical non-equilibrium hydrodynamics in SS was investigated for a long
   time \cite{and,hydro}. These hydrodynamics break down at very low temperature where quantum
   fluctuations dominate. However, the quantum nature of the excitations in
   the SS has not been studied yet. Here,
   we will study the quantum characteristics of low energy excitations in the SS.
   For example, how the phonon spectra in the SS differ from that
   in a NS and how the SF mode in the SS differs from that in
   a SF. The following effective Lagrangian in  the imaginary time describing the low energy excitations inside a SS
   can be seen just from symmetry point of view:
\begin{eqnarray}
   {\cal L}  & = &  \frac{1}{2} [ \rho_{n}  ( \partial_{\tau} u_{\alpha} )^{2} +
    \lambda_{\alpha \beta \gamma \delta} u_{\alpha \beta
    } u_{\gamma \delta } ]    \nonumber  \\
   & +  &  \frac{1}{2}[ \kappa ( \partial_{\tau} \theta )^{2} +
    \rho^{s}_{\alpha \beta} \partial_{\alpha} \theta \partial_{\beta} \theta ]
    + a_{\alpha \beta }  u_{\alpha \beta } i  \partial_{\tau} \theta
\label{ss1}
\end{eqnarray}
    where $ u_{\alpha} = u_{\alpha}( \vec{x},\tau) $ is the lattice displacement,
    the first term is the phonon part, the second term is the
    superfluid part, the last term is the crucial coupling between the
    phonon part and the Berry phase term from the superfluid part which
    comes from $ a_{\alpha \beta} u_{\alpha \beta}
     \psi^{\dagger} \partial_{\tau} \psi $ term in the phase representation of
     the superfluid order parameter $ \psi \sim e^{i \theta} $.
    $ \rho_{n} $ is the normal density, $ u_{\alpha \beta }= \frac{1}{2}( \partial_{\alpha}
    u_{\beta} + \partial_{\beta} u_{\alpha} ) $ is the linearized  strain
    tensor, $ \lambda_{\alpha \beta \gamma \delta} $ are the bare
    elastic constants dictated by the symmetry of the
    lattice, it has 5 (2) independent elastic constants for a hcp  ( isotropic )
    lattice.
    $ \theta $ is the phase and $ \kappa $ is the SF compressibility. For an $ hcp $ crystal, $ \rho^{s}_{\alpha \beta} $
    is the SF stiffness which has the same symmetry as
    $  a_{\alpha \beta }= a_{z} n_{\alpha} n_{\beta} + a_{\perp} (\delta_{\alpha
    \beta}- n_{\alpha} n_{\beta} ) $ with $ \vec{n} $ a unit vector
    points along the preferred axis of the crystal. For an isotropic
    or a cubic crystal $  a_{\alpha \beta }= a \delta_{\alpha
    \beta} $.  In the
    following, we discuss two extreme cases: isotropic solid and $ hcp $ lattice
    separately. Usual samples are between the two extremes, but can be made very close to an  $ hcp $ crystal.

   {\sl (a) Isotropic solid: }
    A truly isotropic solid can only be realized in a highly poly-crystalline
    sample. Usual samples are not completely isotropic. However, we
    expect the simple physics brought about in an isotropic solid
    may also apply qualitatively to other samples which is very
    poly-crystalline.

    For an isotropic solid, $ \lambda_{\alpha \beta \gamma \delta}=
    \lambda \delta_{\alpha \beta} \delta_{\gamma \delta}
    + \mu ( \delta_{\alpha \gamma } \delta_{\beta \delta}+ \delta_{\alpha \delta} \delta_{\beta \gamma
    } ) $ where $ \lambda $ and $ \nu $ are Lame coefficients, $
    \rho^{s}_{\alpha,\beta}= \rho^{s} \delta_{\alpha,\beta}, a_{\alpha,\beta}= a
    \delta_{\alpha,\beta} $. In $ ( \vec{q}, \omega_{n} ) $ space, Eqn.\ref{ss1} becomes:
\begin{eqnarray}
    {\cal L}_{is} & = &  \frac{1}{2}[ \rho_{n} \omega^{2}_{n} + ( \lambda+2 \mu
    ) q^{2} ] |u_{l}( \vec{q},\omega_{n} ) |^{2}         \nonumber  \\
     & + & \frac{1}{2} [ \kappa \omega^{2}_{n} + \rho_{s} q^{2} ] |\theta ( \vec{q},\omega_{n} ) |^{2}
                    \nonumber  \\
     & + &  a q \omega_{n} u_{l}( -\vec{q}, - \omega_{n} ) \theta ( \vec{q},\omega_{n} )
                    \nonumber  \\
    & + &  \frac{1}{2}[ \rho_{n} \omega^{2}_{n} + \mu q^{2} ] |u_{t}( \vec{q},\omega_{n} ) |^{2}
\label{is}
\end{eqnarray}
     where $ u_{l}( \vec{q},\omega_{n} )= i q_{i} u_{i}(
     \vec{q},\omega_{n})/q $ is the longitudinal component,
     $ u_{t}( \vec{q},\omega_{n} )= i \epsilon_{ij} q_{i} u_{j}( \vec{q},\omega_{n}
     )/q $ are transverse components of the
     displacement field. Note that Eqn.\ref{is}
     shows that only longitudinal component couples to the
     superfluid  $ \theta $ mode, while the two transverse components
     are unaffected by the superfluid mode. This is expected, because
     the superfluid mode is a longitudinal density mode itself which
     does  not couple to the transverse modes.

     From Eqn.\ref{is}, we can identify the
     longitudinal-longitudinal phonon correlation function $ \langle u_{l} u_{l} \rangle
     $ and also $ \langle \theta \theta \rangle $ and $ \langle u_{l} \theta \rangle $ correlation
     functions. By doing the
     analytical continuation $ i \omega_{n} \rightarrow \omega + i
     \delta $, we can identify the two poles of  all the correlation
     functions at $ \omega^{2}_{\pm}= v^{2}_{\pm} q^{2} $ where the
     explicit expressions of  the two velocities $ v_{\pm} $ are given in \cite{long},
     but are not needed in our discussions. It is easy to show that $ v_{+} > v_{p}> v_{s} > v_{-} $ and
    $ v^{2}_{+} + v^{2}_{-}  >  v^{2}_{p} + v^{2}_{s} $, but  $ v_{+} v_{-} = v_{p} v_{s} $,
    so $ v_{+} + v_{-} > v_{p} + v_{s} $ ( see Fig.1 ).

    If setting $ a =0 $, then obviously, $ v^{2}_{\pm} $ reduces to the longitudinal phonon
    velocity $ v^{2}_{lp}=  ( \lambda + 2 \mu )/ \rho_{n} $ and
    the superfluid velocity $ v^{2}_{s} = \rho_{s}/\kappa $ respectively.
    Of course, the transverse phonon velocity $ v^{2}_{tp}= \mu /
    \rho_{n} $ is untouched. For notation simplicity, in the
    following, we just use $ v_{p} $ for $ v_{lp} $.
    Inside the SS, due to the very small superfluid density $
    \rho_{s} $, it is expected that $ v_{p} > v_{s} $.
    In fact, in isotropic solid $ He^{4} $, it was measured that $ v_{lp} \sim 450-500
    m/s, v_{t} \sim 230 \sim 320 m/s $ and $ v_{s} \sim 366 m/s $ near the melting curve \cite{melt}.
    The size of the coupling constant $ a $ was estimated to be
    $ \sim 0.1 $  from the slope of the melting curve \cite{elas,dor}.
    So $ v_{+} $ ( $ v_{-} $ ) are about $ 10 \% $ above ( below ) $ v_{p} $ ( $ v_{s} $ ).

\begin{figure}
\includegraphics[width=6cm]{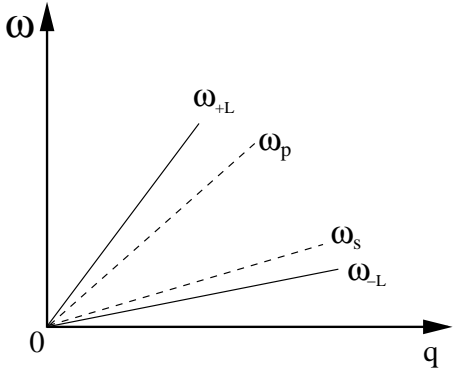}
\caption{ The elementary low energy excitations inside a supersolid.
The coupling between the phonon mode $ \omega_{p} = v_{p}q $ ( the
upper dashed line ) and the superfluid mode $ \omega_{s}=v_{s}q $ (
the lower dashed line ) leads to the two new longitudinal  modes $
\omega_{\pm}=v_{\pm} q $ ( solid lines ) in the SS. The transverse
mode stays the same as that in a normal solid  and  is not shown. }
\label{fig1}
\end{figure}

    The two longitudinal modes in the SS can be understood from an
    intuitive picture: inside the NS, it was argued in \cite{van} that
    there must be a diffusion mode of vacancies in the NS. Inside
    the SS, the vacancies condense and lead to the extra superfluid mode.
    So the diffusion mode in the NS is replaced by the SF  $ \theta $ mode in
    the SS. Its coupling to the lattice phonon modes lead to the
    elementary excitations in a SS shown in Fig.1.

{\sl (b) $ hcp $ crystal: }
  Usual single $ hcp $ crystal samples may also contain dislocations,
  grain boundaries. Here we ignore these line and plane defects
  and assume that there are only vacancies whose condensation leads
  to the superfluid density wave inside the supersolid \cite{qglprl,long}.

  For a uni-axial crystal such as an $ hcp $ lattice, the action is:
\begin{eqnarray}
   {\cal L}_{hcp}  & = &  \frac{1}{2} [ \rho_{n}  ( \partial_{\tau} u_{\alpha} )^{2} +
   K_{11} ( u^{2}_{xx}+u^{2}_{yy} ) + 2 K_{12} u_{xx} u_{yy}
   \nonumber  \\
    &  + &  K_{33} u^{2}_{zz} + 2 K_{13} ( u_{xx}+u_{yy} ) u_{zz}
    \nonumber  \\
    &  + &  2 ( K_{11}-K_{12} ) u^{2}_{xy} + K_{44} ( u^{2}_{yz} +
      u^{2}_{xz} ) ]   \nonumber  \\
   & +  &  \frac{1}{2}[ \kappa ( \partial_{\tau} \theta )^{2} +
    \rho^{s}_{z} ( \partial_{z} \theta )^{2}+  \rho^{s}_{\perp}
    ( ( \partial_{x} \theta )^{2}+ ( \partial_{y} \theta )^{2})   ]
     \nonumber  \\
   & + & [ a_{z} \partial_{z} u_{z} + a_{\perp}
   ( \partial_{x} u_{x} + \partial_{y} u_{y})  ] i  \partial_{\tau} \theta
\label{hcp}
\end{eqnarray}

    If $ \vec{q} $ is along $ \hat{z} $ direction, namely $ q_{z}
    \neq 0, q_{x}=q_{y}=0 $, then Eqn.\ref{hcp}  simplifies to:
\begin{eqnarray}
    {\cal L}^{z}_{hcp} & = &  \frac{1}{2}[ \rho_{n} \omega^{2}_{n} +
    K_{33} q^{2}_{z} ] |u_{z}( q_{z}, \omega_{n} ) |^{2}         \nonumber  \\
     & + & \frac{1}{2} [ \kappa \omega^{2}_{n} + \rho^{s}_{z} q^{2}_{z} ] |\theta ( q_{z},\omega_{n} ) |^{2}
                    \nonumber  \\
     & - & i a_{z} q_{z} \omega_{n} u_{z}( -q_{z}, - \omega_{n} ) \theta ( q_{z},\omega_{n} )
                    \nonumber  \\
    & + &  \frac{1}{2}[ \rho_{n} \omega^{2}_{n} + K_{44} q^{2}_{z}/4 ]|u_{t}( q_{z}, \omega_{n} ) |^{2}
\label{hcpz}
\end{eqnarray}
     where $ |u_{t}( q_{z}, \omega_{n} ) |^{2}= |u_{x}( q_{z},\omega_{n} ) |^{2} +  |u_{y}( q_{z},\omega_{n} ) |^{2} $
     stand for the two transverse modes with the velocity  $ v^{2}_{t}= K_{44}/4 \rho_{n} $. The superfluid mode
     only couples to the longitudinal $ u_{z} $
     mode, while the two transverse modes $ u_{x}, u_{y} $ are decoupled.
     Eqn.\ref{hcpz} is identical to Eqn.\ref{is} after the replacement $ u_{z}
     \rightarrow u_{l}, K_{33} \rightarrow \lambda + 2 \mu, a_{z}
     \rightarrow a $, so $ v^{2}_{lp}=K_{33}/\rho_{n} $. It was found that $ v_{lp} \sim 540 m/s, v_{t}
     \sim 250 m/s $ when $ \vec{q} $ is along the $ \hat{z} $ direction \cite{sound}.
     Fig.1 follows after these replacements.

    Similarly, we can work out the action in the $ xy $ plane where
    $ q_{z}=0, q_{x} \neq 0, q_{y} \neq 0 $.
    Then $ u_{z} $ mode is decoupled, only $ u_{x}, u_{y} $ modes
    are coupled to the superfluid mode:
\begin{eqnarray}
    {\cal L}^{xy}_{hcp} & = &  \frac{1}{2}[
    \rho_{n}  ( \partial_{\tau} u_{\alpha} )^{2}
   + K_{11} ( u^{2}_{xx}+u^{2}_{yy} )
       \nonumber \\
   & + & 2 K_{12} u_{xx} u_{yy} + 2 ( K_{11}-K_{12} ) u^{2}_{xy} ]
     \nonumber  \\
   & +  &  \frac{1}{2}[ \kappa ( \partial_{\tau} \theta )^{2}
    + \rho^{s}_{\perp} ( \partial_{\alpha} \theta )^{2} ]
     \nonumber  \\
   & + &  a_{\perp} \partial_{\alpha} u_{\alpha} i  \partial_{\tau} \theta
      \nonumber  \\
   & + &  \frac{1}{2}[ \rho_{n}  ( \partial_{\tau} u_{z} )^{2}
      + K_{44}/4 ( \partial_{\alpha} u_{z} )^{2}]
\label{hcpxy}
\end{eqnarray}
     where $ \alpha, \beta =x,y $. By comparing Eqn.\ref{hcpxy} with
     Eqn.\ref{is}, we can see that $ K_{11} \rightarrow \lambda + 2 \mu,
     K_{12} \rightarrow \lambda $, so  $ v^{2}_{lp}=K_{11}/\rho_{n} $ and all the discussions in the isotropic case
     can be used here after the replacements.
     Fig.1 follows after these replacements.
     Namely, only the longitudinal component  in
     the $ xy $ plane is coupled to the $ \theta $ mode, while the
     transverse mode in the $ xy $ plane with velocity
     $ v^{2}_{txy}= ( K_{11}-K_{12})/2 \rho_{n} $ is decoupled. Obviously the transverse mode along
     $ \hat{z} $ direction $ u_{z} $ mode with the velocity $ v^{2}_{tz}= K_{44}/4 \rho_{n} $ is also decoupled.
     Note that the two transverse modes have different velocities.
     It was found that $ v_{lp} \sim 455 m/s, v_{tz}
     \sim 255 m/s, v_{txy} \sim 225 m/s $ when $ \vec{q} $ is along the $ xy $ plane \cite{sound}.

     Along any general direction $ \vec{q} $, strictly speaking, one can not even
     define longitudinal and transverse modes, so the general action
     Eqn.\ref{hcp} should be used.
     Despite the much involved $ 4 \times 4 $ matrix
     diagonization in $ u_{x},u_{y},u_{z}, \theta $, we expect the qualitative physics is still
     described by Fig.1.


    Recent inelastic neutron scattering (INS) did not detect any changes in atom kinetic energy
    $ \frac{1}{2} \rho_{n} ( \partial_{\tau} u_{\alpha} )^{2} $ in the temperature
    range $ T=70 \ mK-400 \ mK $  \cite{neukin} and atomic momentum
    distribution function $ n( \vec{k} ) $ with $ n_{0}=0 $  within $ T=80 mK-500 mK
    $ \cite{neubec}. these facts exclude the existence of SS at $ T
    > 70 mK $.  It was known that INS is a very
    powerful tool to measure  the phonon spectra in a normal solid (NS). We
    expect that if SS indeed exists, the INS  can also be used to detect the predicted elementary
    low energy excitation spectra in the SS shown in Fig.1. Namely,
    the INS should be able to map out the dispersion of  two longitudinal modes $
    \omega_{\pm} $ in Fig.1 and the two transverse modes when $ T < T_{SS} $.
    The neutron scattering cross-section from the $ \omega_{\pm L }
    $ and the spectral weight distribution between the two $ \pm L $ modes
    will be calculated \cite{weight}.

{\bf 3. Debye-Waller factor in the X-ray scattering from the SS: }
    It is known that due to zero-point quantum motion in any NS at very low temperature, the X-ray
    scattering amplitude $ I (\vec{G}) $ will be diminished by a Debye-Waller (DW) factor
    $ \sim e^{- \frac{1}{3} G^{2} \langle u^{2}_{\alpha}\rangle } $ where $ u_{\alpha} $
    is the lattice phonon modes in Eqn.\ref{ss1}. In Eqn.\ref{ss1},
    if the coupling between the $ \vec{u} $ and $ \theta $ were
    absent, then the DW factor in the SS would be the same as
    that in the NS.
    By taking the ratio $ I_{SS}( \vec{G})/I_{NS}( \vec{G}) $ at a given reciprocal lattice vector
    $ \vec{G} $, then this DW factor drops
    out. However, due to this coupling,  the $ \langle   u^{2}_{\alpha}\rangle
    $ in SS is different than that in NS, so the DW factor
    will {\sl not} drop out in the ratio.
    In this section, we will calculate this ratio and see how to take care of this factor
    when comparing with the X-ray scattering data.

   The density order parameter at the reciprocal lattice vector $ \vec{G} $ is
  $ \rho_{ \vec{G} }( \vec{x},\tau )  = e^{ i \vec{G} \cdot \vec{u}( \vec{x},\tau )  } $,
  then $ \langle    \rho_{ \vec{G} }( \vec{x},\tau ) \rangle = e^{-\frac{1}{2}
  G_{i}G_{j} \langle    u_{i} u_{j} \rangle } $. The Debye-Waller factor:
\begin{equation}
   I( \vec{G} ) =
  | \langle    \rho_{ \vec{G} }( \vec{x},\tau ) \rangle |^{2}=  e^{- G_{i}G_{j} \langle    u_{i}( \vec{x},\tau ) u_{j}( \vec{x},\tau ) \rangle }
\label{dw}
\end{equation}
  where the phonon-phonon  correlation  function is:
\begin{equation}
   \langle    u_{i} u_{j} \rangle =  \langle   u_{l}u_{l} \rangle  \hat{q}_{i}\hat{q}_{j} +  \langle   u_{t}u_{t} \rangle ( \delta_{ij}- \hat{q}_{i}\hat{q}_{j} )
\label{uu}
\end{equation}
   where $ \hat{q}_{i}\hat{q}_{j}= \frac{ q_{i} q_{j} }{ q^{2}} $.  .

   Then substituting Eqn.\ref{uu} into Eqn.\ref{dw} leads to:
\begin{equation}
  \alpha( \vec{G} ) = I_{SS}( \vec{G} )/I_{NS}(
  \vec{G})= e^{-\frac{1}{3} G^{2} ( \Delta u^{2} )_{l} }
\end{equation}
  where $  ( \Delta u^{2} )_{l}= \langle   u^{2}_{l}( \vec{x},\tau ) \rangle_{SS}- \langle   u^{2}_{l}( \vec{x},\tau ) \rangle_{NS} $ and
  the transverse mode drops out,  because it stays the same in the SS and in the NS.

    From Eqn.\ref{is}, it is easy to see that $ ( \Delta u^{2} )_{l} < 0   $, namely,
    the longitudinal vibration amplitude in SS is {\em smaller} that that
    in NS. The $ \alpha( \vec{G} )( T )= e^{-\frac{1}{3} G^{2}( \Delta
    u^{2} )_{l}} > 1 $. This is expected, because the SS state is the ground
    state at $ T < T_{SS} $, so the longitudinal vibration amplitude
    should be reduced compared to the corresponding NS with the same
    parameters $ \rho_{n}, \lambda, \mu $. It is easy to show that
    $ ( \Delta u^{2} )_{l}( T=0) < 0 $ and $  ( \Delta u^{2} )_{l}( T )-( \Delta u^{2} )_{l}(
    T=0) \sim T^{2} >0 $.
    Namely, the difference in the ratio  will {\em decrease } as $ T^{2} $ as the
    temperature increases. Of course, when $ T $ approaches $ T_{SS} $ from below, the
    difference vanishes, the $ \alpha( \vec{G} ) $ will approach $ 1 $ from above, the SS turns into a NS.
    The density-density correlation function was studied in
    \cite{long}. Unfortunately, very recent INS \cite{neudw} did not detect
    the predicted anomaly in the Debye-Waller factor within $ T=140 mK -1K $. This fact
    indicates the absence of SS when $ T > 140 mK $.


{\bf 4.  Vortex loops in supersolid }
   In section 3, we studied the low energy excitations shown in the
   Fig.1 by neglecting the  topological vortex loop in the phase $ \theta $. Here, we will
   study how the vortex loop interaction in the SS differ from that in
   the SF.   For simplicity,
   in the following, we only focus on the isotropic case. The
   formulations can be generalized to the $ hcp $ case
   straightforwardly. We can perform a duality transformation on
   Eqn.\ref{ss1} to the vortex loop representation in terms of the 6
   components anti-symmetric tensor gauge field $  a_{\mu \nu}=- a_{\nu \mu} $
   and the 6 components  anti-symmetric tensor vortex
   loop current $ j^{v}_{\mu \nu}= \frac{1}{2 \pi}  \epsilon_{\mu \nu \lambda \sigma
   } \partial_{\lambda } \partial_{\sigma } \theta $  due to the topological phase winding in $ \theta $.
    It is the most convenient to choose the Coulomb gauge $ \partial_{\alpha} a_{\alpha
    \beta}=0 $ to get rid of the longitudinal component, then
    the transverse component is $ a_{t}= i \epsilon_{\alpha \beta
    \gamma} q_{\alpha} a_{\beta \gamma} /q $. It can be shown that $
    |a_{t}|^{2}=2 | a_{\alpha \beta} |^{2} $.
    Then Eqn.\ref{ss1} is dual to:
\begin{eqnarray}
 {\cal L}_{v} & = &  \frac{1}{2}[ \rho_{n} \omega^{2}_{n} + ( \lambda+2 \mu
     + a^{2}/\kappa ) q^{2} ] |u_{l}( \vec{q},\omega_{n} ) |^{2}        \nonumber  \\
     & + &  \frac{1}{2} ( q^{2}/\kappa + \omega^{2}_{n}/\rho_{s} ) |a_{t}|^{2}
     + \frac{2}{ \rho_{s} } q^{2} | a_{0 \alpha} |^{2}
                    \nonumber  \\
     & - &  a q^{2}/\kappa  u_{l}( -\vec{q}, - \omega_{n} ) a_{t}( \vec{q},\omega_{n} )
                    \nonumber  \\
    & + & i 2 \pi j^{v}_{ 0 \alpha } a_{0 \alpha} + i 2 \pi
    j^{v}_{\alpha \beta} a_{\alpha \beta }
\label{prop}
\end{eqnarray}
    where the transverse phonon mode $ u_{t} $ was omitted, because
    it stays the same as in the NS as shown in Eqn.\ref{is}.

   It is easy to see that only $ a_{t} $ has the dynamics, while $ a_{0
   \alpha} $ is static. This is expected, because although $ a_{\mu
   \nu} $ has 6 non-vanishing components, only the transverse component $ a_{t}
   $ has the dynamics which leads to the original gapless superfluid
   mode $ \omega^{2}= v^{2}_{s} q^{2} $. Eqn.\ref{prop} shows that
   the coupling is between the longitudinal phonon mode $ u_{l} $ and the
   transverse gauge mode $ a_{t} $. The vortex loop density is
   $ j^{v}_{ 0 \alpha }= \frac{ 1}{2 \pi} \epsilon_{\alpha \beta
    \gamma} \partial_{\beta}   \partial_{\gamma} \theta $ and the
    vortex current is $ j^{v}_{ \alpha  \beta }= \frac{ 1}{2 \pi} \epsilon_{\alpha \beta
    \gamma} [ \partial_{0},  \partial_{\gamma}] \theta $. Integrating out the $ a_{0 \alpha} $, we
    get the vortex loop density-density interaction:
\begin{equation}
    \pi \rho_{s} \int^{\beta}_{0} d \tau \int dx dy j^{v}_{ 0 \alpha
    }( \vec{x}, \tau ) \frac{1}{|x-y|} j^{v}_{ 0 \alpha }( \vec{y}, \tau )
\end{equation}
     Namely, the vortex loop density- density interaction in SS stays  as $ 1/r $ which is the
     same as that in SF. Therefore, a single vortex loop energy
     and the critical transition temperature $ T_{3dxy} $
     \cite{long} is solely determined by the superfluid density $ \rho_{s} $
     independent of any other parameters in Eqn.\ref{is},
     except that the vortex core of the vortex loop is much larger
     than that in a superfluid \cite{qglprl}.
     The critical behaviors of the vortex loops close to the 3d XY
     transition was studied in \cite{loop}.

     Integrating out the $ a_{t} $,
     we get the vortex loop current- current interaction:
\begin{equation}
    2 \pi^{2} j^{v}_{\alpha \beta} (-\vec{q},-\omega_{n} )
    D_{\alpha \beta, \gamma \delta} ( \vec{q}, \omega_{n} )  j^{v}_{\gamma \delta }(\vec{q},\omega_{n} )
\end{equation}
     where $ D_{\alpha \beta, \gamma \delta} ( \vec{q}, \omega_{n} )
     =( \delta_{\alpha \gamma} \delta_{\beta \delta}- \delta_{\beta \gamma} \delta_{\alpha \delta}
     - \frac{ q_{\beta} q_{\delta} }{q^{2}} \delta_{\alpha \gamma}
     - \frac{ q_{\alpha} q_{\gamma} }{q^{2}} \delta_{\beta \delta}
     + \frac{ q_{\alpha} q_{\delta} }{q^{2}} \delta_{\alpha \delta}
     + \frac{ q_{\beta} q_{\gamma} }{q^{2}} \delta_{\alpha \delta }
     )  D_{t}(\vec{q}, \omega_{n} ) $ where  $ D_{t}(\vec{q}, \omega_{n}
     ) $ is the $ a_{t} $ propagator. Defining
      $  \Delta D_{t}(\vec{q}, \omega_{n}) = D^{SS}_{t}(\vec{q}, \omega_{n})-D^{SF}_{t}(\vec{q}, \omega_{n}) $
      as the difference between the $a_{t} $ propagator in the SS and the
      SF. For simplicity, we just give the expression for the equal
      time $ \Delta D_{t}( \vec{x}-\vec{x}^{\prime},\tau=0 )=
       \frac{c}{ (  \vec{x}-\vec{x}^{\prime}  )^{2} } $ where $ c $
       is a positive constant if assuming SS has the same parameters $ \kappa, \rho_{s}
       $ as the SF.

{\bf 5. Specific heat in the SS }
     It is well known that  at low $ T $, the specific heat in the NS is $
     C^{NS}= C^{NS}_{lp} + C^{NS}_{tp} + C_{van} $
     where $ C^{NS}_{lp} =
     \frac{ 2 \pi^{2}}{15} k_{B} ( \frac{k_{B}T}{\hbar v_{lp} })^{3}
     $  is from the longitudinal phonon mode and $ C^{NS}_{tp} =
     2 \times \frac{ 2 \pi^{2}}{15} k_{B} ( \frac{k_{B}T}{\hbar v_{tp} })^{3} $
     is from the two transverse phonon modes, while $ C_{van} $ is
     from the vacancy contribution. $ C_{van} $ was calculated in
     \cite{heat} by assuming 3 different kinds of models for the
     vacancies. So far, there is no consistency between the calculated $ C_{van}
     $ and the experimentally measured one \cite{heat,ander}.
     The specific heat in the SF $ C^{SF}_{v}= \frac{ 2 \pi^{2}}{15} k_{B} ( \frac{k_{B}T}{\hbar v_{s} })^{3} $
     is due to the SF mode $ \theta $.  From Eqn.\ref{is}, we can find the specific heat in the SS:
\begin{equation}
 C^{SS}_{v}= \frac{ 2 \pi^{2}}{15} k_{B} ( \frac{k_{B}T}{\hbar
v_{+} })^{3} + \frac{ 2 \pi^{2}}{15} k_{B} ( \frac{k_{B}T}{\hbar
v_{-} })^{3} + C^{tp}
 \label{spec}
\end{equation}
    where $ C^{tp} $ stands for the contributions from the
    transverse phonons which are the same as those in the NS.

     It was argued in \cite{qglprl}, the critical regime of finite temperature  NS to
     SS transition in Fig.1 is much narrower than the that of SF to
     the NL transition, so there should be a jump in the specific
     heat at $ T=T_{SS} $. From Eqn.\ref{spec}, it is easy to see that
     the excess entropy due to the vacancies is:
\begin{equation}
 \Delta S = \int^{T_{SS}}_{0} dT \frac{C_{van}}{T} = \frac{ 2 \pi^{2}}{45} k_{B} ( \frac{k_{B} T_{SS}
 }{\hbar})^{3}( \frac{1}{ v^{3}_{+} }+ \frac{1}{ v^{3}_{-} }- \frac{1}{ v^{3}_{lp} })
 \label{enss}
\end{equation}
       where  $ \Delta S >0 $ is dominated by the lower branch $  v_{-} < v_{lp} $ in Fig.1.
       Using the molar volume $ v_{0} \sim 20 cm^3/mole $ of solid $^{4} He $ and $ T_{ss} \sim 100 mK $,
       we can estimate
       the $ \Delta S $ per mole is $ \sim 10^{-5} R $ where $ R $
       is th gas constant. This estimate is 3 orders magnitude
       smaller than that in \cite{entropy,melt} where the SS state was
       taken simply as the boson condensation of non-interacting
       vacancies. Our estimate is indeed consistent with recent
       experiment on specific heat \cite{heatpsu}.

{\bf 6. Conclusions: }
     In this paper, starting from the quantum Ginsburg-Landau theory
     developed in \cite{qglprl,long}, we worked out the elementary excitations inside a supersolid. We
     found that the elementary excitations have two longitudinal modes $
     \omega_{\pm}=v_{\pm} q $ shown in Fig.1. The transverse modes
     in the SS stays the same as those in the NS.
     The $ \omega_{\pm} $ are estimated to be  $ 10 \% $ higher ( lower ) than the sound speed
     in the normal solid and  the superfluid respectively.
      Then we calculated the experimental
     signature of the two modes. We found that the longitudinal
     vibration in the SS is smaller than that in the NS ( with the same corresponding solid parameters ), so the DW
     factor at a given reciprocal lattice vector is larger than
     that in the NS. The density-density correlation function in the
     SS is weaker than that in the NS. By going to the dual
     vortex loop representation, we found the vortex loop
     density-density interaction in a SS stays the same as that in the
     SF ( with the same corresponding superfluid parameters ), so
     the vortex loop energy and the corresponding SS to NS transition
     temperature is solely determined by the superfluid density and
     independent of any other parameters. The vortex current-current
     interaction in a SS is stronger than that in the SF. The specific heat
     in the SS is given by the sum from the transverse phonons and the two longitudinal
     phonons and still shows $ \sim T^{3} $ behavior.
     The excess entropy due to the vacancies was estimated to be 3 order of magnitude smaller
     than the previous idea bose gas estimation. Comparison with
     very recent neutron scattering measurements are made. No matter
     if the SS exists in $ ^{4}He $, the results achieved should be
     interesting in its own right and may have applications in other
     systems such as the possible excitonic supersolid in
     electron-hole bilayer systems \cite{ehbl}.

     The research at KITP was supported in part by the NSF under
     grant No. PHY-05-51164 and at KITP-C
     by the Project of Knowledge Innovation Program (PKIP) of Chinese Academy of Sciences.

\end{document}